\documentclass[conference]{IEEEtran}
\usepackage{graphicx,subfigure,psfrag,epsfig,epsf,latexsym,hhline,amsmath,amssymb,amsthm,multirow}

\IEEEoverridecommandlockouts
\begin{document}

\title{Multilevel Coding Schemes for Compute-and-Forward\thanks{This work was supported by the National Science Foundation under Grant CCF 0729210}}
\author{Brett Hern and Krishna Narayanan\\
Texas A\&M University\\
College Station\\
TX 77843}
\maketitle

\begin{abstract}
We consider the design of coding schemes for the wireless two-way relaying channel when there is no channel state information at the transmitter. In the spirit of the compute and forward paradigm, we present a multilevel coding scheme that permits the recovery of a class of functions at the relay. We define such a class of functions and derive rates that are universally achievable over a set of channel gains when this class of functions is used at the relay. We develop our framework with general modulation formats in mind, but numerical results are presented for the case where each node transmits using the QPSK constellation. Numerical results with QPSK show that substantially higher rates are achievable with our proposed approach than those achievable by always using a fixed function or adapting the function at the relay but coding over GF(4).
\end{abstract}


\section{Introduction}
\begin{figure}
    \centering
    \includegraphics[width=2.5in]{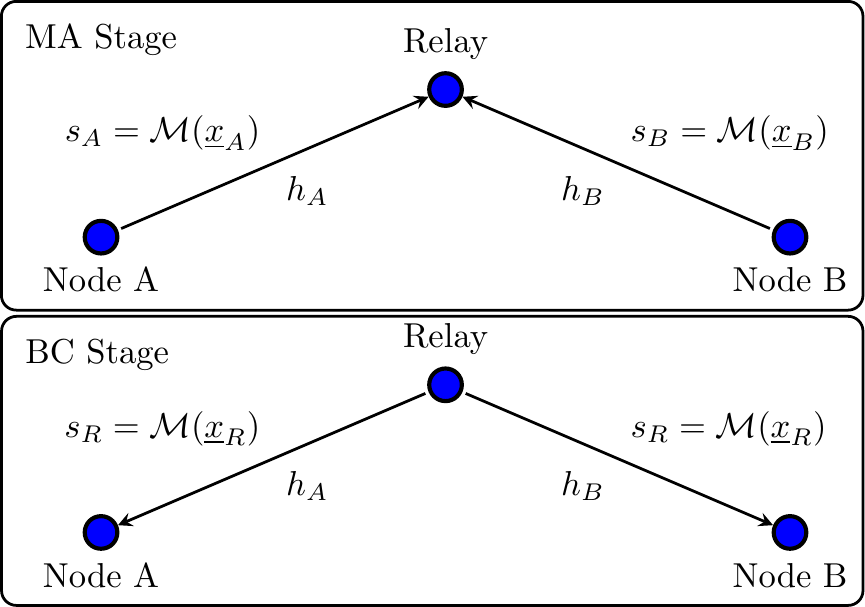}
    \caption{Diagram of two-way relaying channel with PLNC.}
    \label{fig:RelayChannel}
\end{figure}

Physical layer network coding (PLNC) or Compute and Forward is a new paradigm in wireless networks where each relay in a network decodes a function of the transmitted messages and broadcasts the value of this function to the other nodes in the network. This has been shown to provide significant increase in achievable rates for some networking problems \cite{popovski2009coded}, \cite{DBLP:journals/corr/abs-0805-0012}, \cite{nazer2007lattice}.
%
%
An example of such a problem where compute and forward has been shown to be effective is the two-way relaying problem shown in Fig. \ref{fig:RelayChannel}. Node $A$ has data to send to node $B$ and vice versa. The relay $R$ is included to assist in this communication, and it is assumed that there is no direct link between nodes $A$ and $B$. Near optimal coding schemes have been designed for this problem for the case where there is no fading in the channel in \cite{DBLP:journals/corr/abs-0805-0012}, \cite{nazer2007lattice}, \cite{nam5capacity} and for the case when there is fading but the transmitter has perfect channel state information in \cite{wilson2009power}.


In this paper, the complex channel coefficients $h_A$ and $h_B$ are assumed to be perfectly estimated at each receiver but unknown to each transmitter. For this scenario, the authors in \cite{koike2009optimized} introduce a scheme called denoise-and-forward which uses channel dependent denoising functions at the relay to minimize the symbol error probability. The relay chooses denoising functions so that the distance profile for constellation points with different labels is optimized. This improves the symbol error rate for transmissions between nodes A and B, however,
denoising is performed purely at the symbol level. There is no natural extension to include error correction at the relay.

Recently, a scheme called compute-and-forward, which allows both adaptation of decoding functions and error correction at the relay has been presented in \cite{nazer2009compute}. In this scheme, the relay decodes an integer combination of the transmitted codewords, where the integer combination is adapted according to the channel gains. They show that such a scheme can be implemented using nested lattice codes to take advantage of the duality between modulus arithmetic on prime order fields and the modular operations of lattice decoding. Their scheme requires the construction of infinite dimensional lattices which is not practical.
The results in \cite{nazer2009compute} are extended in a remarkable way in \cite{feng2010algebraic}, where an algebraic framework is provided to design lattices over principal ideal domains. However, their proposed coding scheme is also based on large dimensional lattice codes.

In this paper, we propose a compute and forward scheme based on multilevel coding (MLC). Unlike the coding schemes in \cite{nazer2009compute}, \cite{feng2010algebraic},  our proposed scheme does not result in a lattice code and uses only linear codes over prime fields (for example, binary linear codes) and, is hence, practical. Yet, it facilitates error correction for a larger class of decoding functions than those proposed in \cite{nazer2009compute}. This is because the class of functions for our scheme is derived from the large set of non-singular square matrices over $\mathbb{F}_p$ in place of the set of non-zero elements in large prime order fields. To the best of our knowledge, such an idea of using multilevel coding and exploiting the linearity over the prime field to adaptively decode linear functions of transmitted codewords is new. Another important contribution in this paper is that our proof for the achievability of rates with the proposed multilevel coding scheme requires a non-trivial extension of the proof of achievability of rates for multilevel coding for the point to point case.


This paper is organized as follows. The key elements of the problem are outlined in Section \ref{sec:ProblemDescription}. Our proposed solution is detailed in Section \ref{sec:ProposedScheme}. The rate which can be achieved by our scheme for the MA stage is given in Section \ref{sec:InfoAnalysis}. These rates are numerically determined for an example where nodes A and B transmit using a QPSK constellation in Section \ref{sec:NumericalResults}. Key results are reiterated in Section \ref{sec:ConcludingRemarks}.


\section{Problem Description} \label{sec:ProblemDescription}
Each node in the relay network is assumed to be half-duplex, so communication is split into two stages, a multiple access (MA) stage and a broadcast (BC) stage. We assume perfect synchronization between the transmitters and mainly focus on the MA stage in this paper.

\subsection{Multiple Access Stage}
Nodes A and B each encode their binary messages $\underline{u}_A$ and $\underline{u}_B$ into codewords $\underline{v}_A\in\mathcal{C}_A$ and $\underline{v}_B\in\mathcal{C}_B$ where $\mathcal{C}_A$ and $\mathcal{C}_B$ are the codebooks used at the nodes $A$ and $B$ respectively. These codewords are mapped to sequences of symbols $\underline{s}_A,\underline{s}_B\in\mathcal{Q}^N$ with $|\mathcal{Q}|=2^\ell$. The relay receives noisy observations of the sum of these symbol sequences according to
\begin{equation}
\underline{y}_R = h_A\underline{s}_A+h_B\underline{s}_B+\underline{w}_R
\end{equation}
where $h_A$ and $h_B$ are complex fading coefficients, and $\underline{w}_R$ is complex additive white Gaussian noise (AWGN) with variance $N_0$.

\subsection{Adaptive Decoding at the Relay}
The main idea proposed in this paper is the construction of a coding scheme such that the relay can reliably decode some function of $\underline{v}_A$ and $\underline{v}_B$ for a desired set of channel conditions $\mathcal{H}\subset\mathbb{C}^2$. Specifically, we jointly design codes $\mathcal{C}_A$ and $\mathcal{C}_B$ and a set of decoding functions $\mathcal{F}$ such that, for any $(h_A,h_B)\in\mathcal{H}$, there exists $f\in\mathcal{F}$ such that the relay can reliably decode $f(\underline{v}_A,\underline{v}_B)$ from $\underline{y}_R$. We require that, given the output of $f(\underline{v}_A,\underline{v}_B)$, node A (B) must be able to unambiguously decode $\underline{v}_B~(\underline{v}_A)$ using its knowledge of $\underline{v}_A~(\underline{v}_B)$. For a given $f\in\mathcal{F}$, we will define an induced codebook at the relay as the codebook corresponding to $f$ i.e.
\begin{equation}
\mathcal{C}_{f,R} = \{f(\underline{v}_A,\underline{v}_B)| \underline{v}_A \in \mathcal{C}_A, \underline{v}_B \in \mathcal{C}_B \}
\end{equation}
It is important to understand the structure of $\mathcal{C}_{f,R}$ since the probability of error in decoding $f(\underline{v}_A,\underline{v}_B)$ from $\underline{y}_R$ depends on $h_A$, $h_B$, and $\mathcal{C}_{f,R}$. The main advantage of our proposed scheme is that it guarantees that choosing one codebook $\mathcal{C}_A$ and $\mathcal{C}_B$ at the transmitter can result in a good induced codebook
$\mathcal{C}_{f,R}$ for a class of functions $\mathcal{F}$. More specifically, it guarantees $\mathcal{C}_{f,R}$ is a member of the ensemble of random coset codes which is an optimal ensemble for achieving the uniform input information rate for the equivalent channel between $f(\underline{v}_A,\underline{v}_B)$ and $\underline{y}_R$ for all $f \in \mathcal{F}$.

The broadcast stage is fairly standard and is identical to that considered in \cite{DBLP:journals/corr/abs-0805-0012}, \cite{nazer2007lattice}.


\section{Proposed Scheme} \label{sec:ProposedScheme}

\subsection{Multilevel Encoder}
\begin{figure}
    \centering
    \includegraphics[width=2.75in]{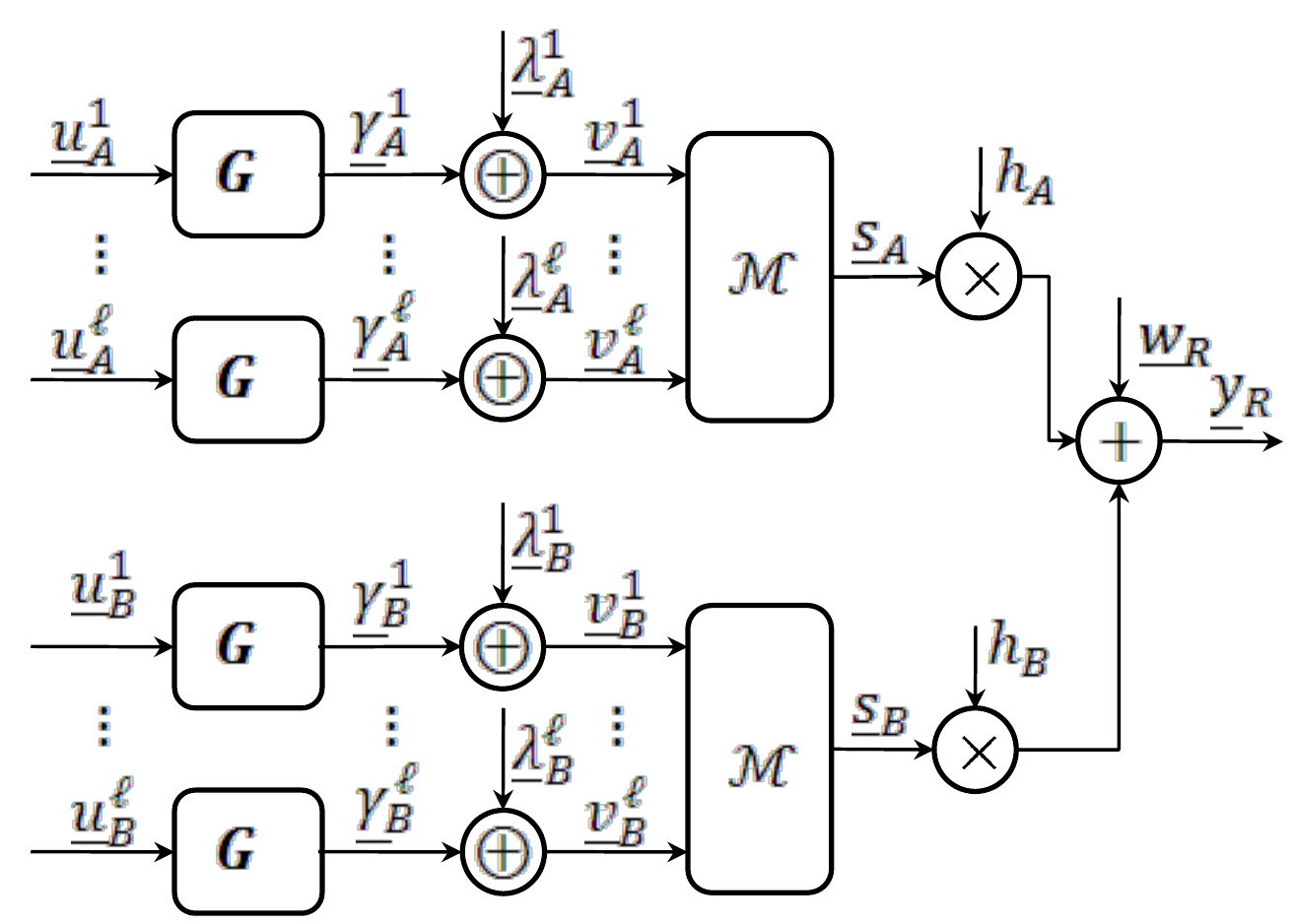}
    \caption{Block Diagram of MLC Coset Encoders for MA Stage.}

    \label{fig:MLCCosetEnc}
\end{figure}

The system model for the multilevel encoder for nodes A and B and the channel model for the MA stage is shown in Fig. \ref{fig:MLCCosetEnc}. The encoder at nodes A and B uses MLC with a different coset of the same linear code $\mathcal{C}$ used at each bit level. For a detailed description of MLC and achievable rates for the point to point channel, see \cite{wachsmann1999multilevel}.

The encoder is described as it pertains to node A to simplify notation. First, the message $\underline{u}_A$ is split into sub-vectors $\underline{u}_A^1,...,\underline{u}_A^\ell$ which form rows of an $\ell\times K$ matrix
\begin{equation}
\mathbf{U}_A = \left[\begin{array}{c} \underline{u}_A^1 \\ \vdots \\ \underline{u}_A^\ell \end{array} \right].
\end{equation}
Each $\underline{u}_A^k,~\{1,...,\ell\}$ is encoded with a linear code $\mathcal{C}$ with generator matrix $\mathbf{G}$ to get codewords $\underline{\gamma}_A^1,...,\underline{\gamma}_A^\ell$. These codewords from the rows of an $\ell\times N$ matrix $\mathbf{\Gamma}_A=\mathbf{U}_A \mathbf{G}$.
Finally, a random binary vector $\underline{\lambda}_A^k$ is added to each $\underline{\gamma}_A^k$. Since each $\underline{\lambda}_A^k$ can be thought of as coset leaders of a random coset of the original linear code. We obtain a codeword of a random coset given by $\underline{v}_A^k=\underline{\lambda}_A^k\oplus\underline{\gamma}_A^k,~k\in\{1,..,\ell\}$. The random coset leaders form an $\ell\times N$ matrix $\mathbf{\Lambda}_A$.
The resulting coset codewords $\underline{v}_A^k$ form the rows of a binary $\ell\times N$ matrix $\mathbf{X}_A$ given by
\begin{equation}
\mathbf{X}_A = \mathbf{U}_A \mathbf{G} \oplus \mathbf{\Lambda}_A = \left[\begin{array}{c} \underline{v}_A^1 \\ \vdots \\ \underline{v}_A^\ell \end{array} \right] = \left[\underline{x}_A[1],...,\underline{x}_A[N]\right].
\end{equation}
Thus each code $\mathcal{C}_A^k,~k\in\{1,...,\ell\}$ will be a different coset of $\mathcal{C}$. The $k_{th}$ row $\underline{v}_A^k$ of $\mathbf{X}_A$ is then a codeword of $\mathcal{C}_A^k$, and the $n_{th}$ column $\underline{x}_A[n]$ is the binary address vector of the $n_{th}$ symbol of $\underline{s}_A$. The $n_{th}$ binary address vector $\underline{x}_A[n]\in\mathbb{F}_2^\ell$ maps to a symbol $\underline{s}_A[n]\in\mathcal{Q}$ through the use of a symbol mapping function $\mathcal{M}:\mathbb{F}_2^\ell\rightarrow\mathcal{Q}$. It should be mentioned here that much of the intuition about the main result in the paper is best obtained by ignoring the fact that cosets are used at each layer and simply considering the use of identical linear codes at each level in the MLC scheme. The coset matrix $\mathbf{\Lambda}_A$ is included to symmetrize the effective channel at the relay (i.e. $\mathbf{\Lambda}_A$ is necessary for the proofs to be correct).

\subsection{Adaptive Decoding at the Relay}
As mentioned previously, the goal of the proposed scheme is to allow the relay to decode a function of the transmitted codewords.
If nodes A and B encode their messages as described, the set of decoding functions $\mathcal{F}$ which the relay can use for decoding is defined as follows.

Define $\mathcal{D}$ as the set of $\ell\times\ell$ binary matrices which are invertible using operations over $\mathbb{F}_2$. The set of functions we consider is given by
\begin{align} \label{eq:FSetDef}
{\mathcal F} &= \{f:\mathbb{F}_{2}^\ell\times\mathbb{F}_{2}^\ell\rightarrow\mathbb{F}_{2}^\ell \nonumber \\
 &~|~ f(\underline{x}_A,\underline{x}_B) = [\mathbf{D}_A \mathbf{D}_B] \left[\begin{array}{c}\underline{x}_A \\ \underline{x}_B \end{array}\right], ~ \mathbf{D}_A,\mathbf{D}_B\in\mathcal{D} \}.
\end{align}
Therefore a given $f\in\mathcal{F}$ is defined by some $\mathbf{D}_A,\mathbf{D}_B\in\mathcal{D}$ from which the relay should attempt to decode a matrix $\mathbf{X}_{f,R}$ given by $\mathbf{X}_{f,R} = [\mathbf{D}_A \mathbf{D}_B] \left[ \begin{array}{c} \mathbf{X}_A \\ \mathbf{X}_B \end{array} \right]$.

Due to the linearity of $[\mathbf{D}_A,\mathbf{D}_B]$ and $\mathbf{G}$, we can express the desired matrix $\mathbf{X}_{f,R}$ as
\begin{align} \label{eq:XrEquiv}
\mathbf{X}_{f,R} &= [\mathbf{D}_A \mathbf{D}_B] \left[ \begin{array}{c} \mathbf{X}_{A} \\  \mathbf{X}_{B} \end{array} \right] = [\mathbf{D}_A \mathbf{D}_B] \left[ \begin{array}{c} \mathbf{U}_{A} \mathbf{G} \oplus \mathbf{\Lambda}_A \\  \mathbf{U}_{B} \mathbf{G} \oplus \mathbf{\Lambda}_B \end{array} \right] \nonumber \\
&= [\mathbf{D}_A \mathbf{D}_B] \left[ \begin{array}{c} \mathbf{U}_{A} \\  \mathbf{U}_{B} \end{array} \right] \mathbf{G} \oplus [\mathbf{D}_A \mathbf{D}_B] \left[ \begin{array}{c} \mathbf{\Lambda}_A \\  \mathbf{\Lambda}_B \end{array} \right] \nonumber  \\
&= \mathbf{U}_{f,R} \mathbf{G} \oplus \mathbf{\Lambda}_{f,R}.
\end{align}
Here, we see that the matrix $\mathbf{X}_{f,R}$ can be written in terms of an effective message $\mathbf{U}_{f,R}$ and coset matrix $\mathbf{\Lambda}_{f,R}$ which can be computed separately based on $f$. Thus the rows of $\mathbf{X}_{f,R}$ are codewords from a different coset code of $\mathcal{C}$. Note that $f$ is applied elementwise to the sequences $\underline{s}_A$ and $\underline{s}_B$.

\begin{figure}
    \centering
    \includegraphics[width=3.5in]{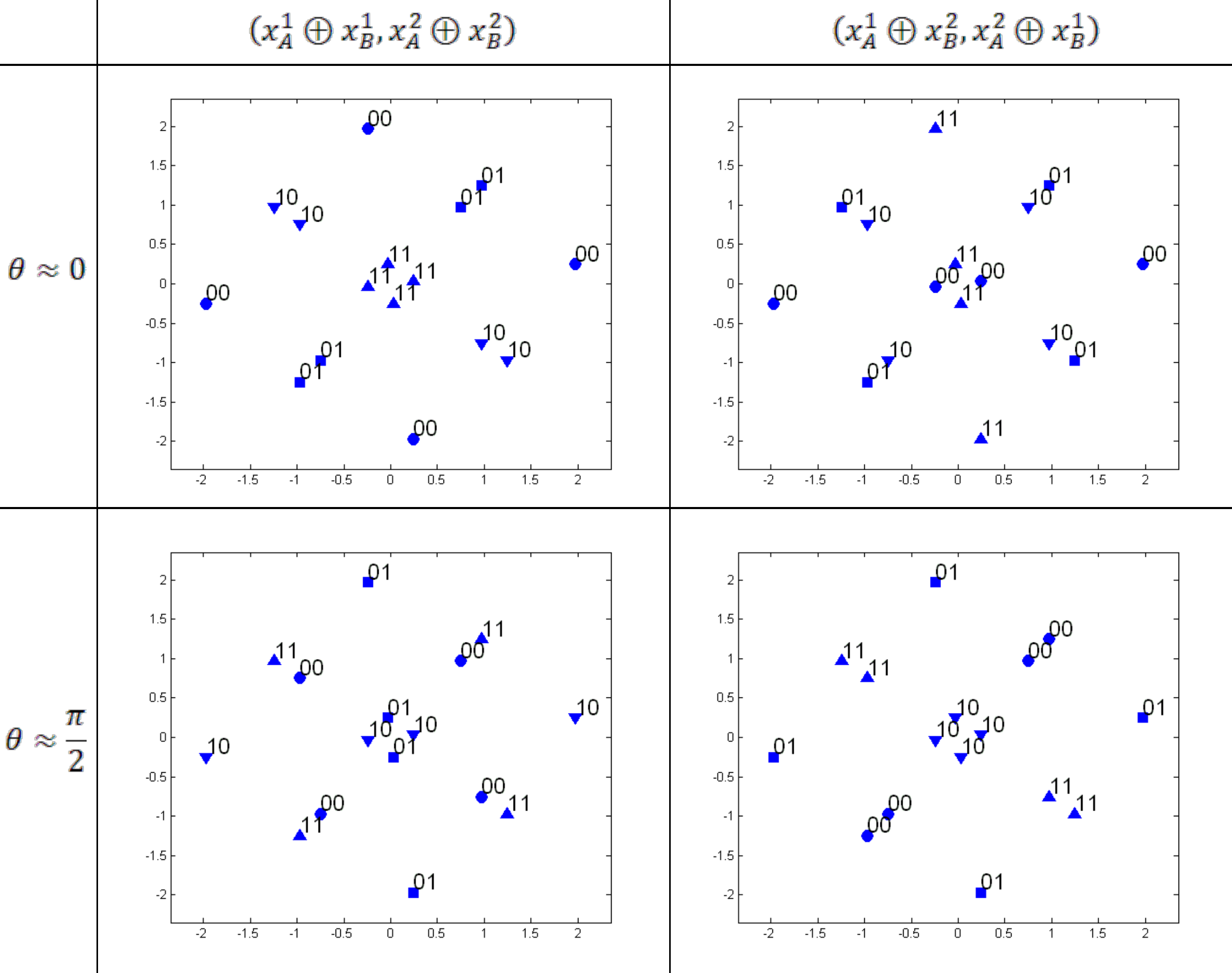}
    \caption{Effective constellation at relay for different values of $\theta$.}

    \label{fig:ConstPlot}
\end{figure}

To illustrate the importance of choosing the decoding function $f$ depending on $(h_A,h_B)$, consider an example with $\mathcal{Q}=\{1,j,-1,-j\}=\{\mathcal{M}(00),\mathcal{M}(01),\mathcal{M}(11),\mathcal{M}(10)\}$ (i.e. QPSK with Gray Labeling).  Further, let $h_A = e^{j \theta_A}$, $h_B = e^{j \theta_B}$, and $\theta = \theta_A -\theta_B$ be the phase difference. Consider the decoding functions $f_1(\underline{x}_A,\underline{x}_B)=[x_A^1\oplus x_B^1,x_A^2\oplus x_B^2],~f_2(\underline{x}_A,\underline{x}_B)=[x_A^1\oplus x_B^2,x_A^2\oplus x_B^1].$
The resulting constellation $\mathcal{Q}_R$ at the relay is shown for different values of $\theta$ in Fig. \ref{fig:ConstPlot}. Note that the complex coordinates of the constellation points are exactly the same, but their labels are different based on $\theta$ and $f\in\{f_1,f_2\}$. When $\theta\approx 0$, $f_1$ appears to have better performance than $f_2$ in terms of the distances between points with unequal labels. The situation is reversed when $\theta\approx \frac{\pi}{2}$. This shows that the performance for a fixed decoding function can vary widely with $\theta$ even when both $|h_A|$ and $|h_B|$ are large.


In order for nodes A and B to be able to unambiguously decode their desired messages, the authors in \cite{koike2009optimized} show that $f$ must satisfy
\begin{align} \label{eq:unambiguous}
&f(\underline{x}_A,\underline{x}_B) \neq f(\underline{x}_A^\prime ,\underline{x}_B) ~ \forall ~ \underline{x}_A \neq \underline{x}_A^\prime  \textrm{~and~} \underline{x}_B \nonumber \\
&f(\underline{x}_A,\underline{x}_B) \neq f(\underline{x}_A,\underline{x}_B^\prime) ~ \forall ~ \underline{x}_B \neq \underline{x}_B^\prime \textrm{~and~} \underline{x}_A.
\end{align}
We call functions that satisfy this property unambiguous.

\emph{Lemma 1:}
For any $\mathbf{D}_A,\mathbf{D}_B \in \mathbf{\mathcal{D}}$, a decoding function $f(\underline{x}_A,\underline{x}_B) = [\mathbf{D}_A \mathbf{D}_B] \left[\begin{array}{c}\underline{x}_A \\ \underline{x}_B \end{array}\right]$ is unambiguous.

\begin{proof}
The full proof is omitted for space, but follows from the invertibility of $\mathbf{D}_A$ and $\mathbf{D}_B$.
\end{proof}

\section{Achievable Information Rates} \label{sec:InfoAnalysis}

\subsection{Achievable Rate for a Given Function}
For a given $f$ and fixed channel gains $h_A$ and $h_B$ the achievable rate region is given by the following theorem. This theorem is the key contribution of this paper.

\emph{Theorem 1:}  Choose some fixed $\mathbf{D}_A,\mathbf{D}_B \in \mathcal{D}$ and define
\begin{equation}
\underline{x}_{f,R} = f(\underline{x}_A,\underline{x}_B) = [\mathbf{D}_A \mathbf{D}_B] \left[\begin{array}{c}\underline{x}_A \\ \underline{x}_B \end{array}\right].
\end{equation}
Choose a subset $\mathcal{S}\subseteq\{1,...,\ell\}$ and define $\overline{\mathcal{S}}=\{1,...,\ell\}\setminus\mathcal{S}$. Divide $\mathcal{S}$ into $p$ non-empty disjoint subsets $\mathcal{S}_1,...,\mathcal{S}_p$ so that $\bigcup_{i=1}^p\mathcal{S}_i = \mathcal{S}$. Let $Z_i,~i\in\{1,...,p\}$ define $p$ i.i.d. Bernoulli random variables with parameter $\frac{1}{2}$. At last, let each row of $\mathbf{X}_A$ and $\mathbf{X}_B$ be encoded using a different coset of the same linear code $\mathcal{C}$. Then there exists a linear code $\mathcal{C}$ of rate $\mathcal{R}$ for which the relay can reliably decode $\mathbf{X}_{f,R}$ as long as $\mathcal{R}$ satisfies
\begin{align} \label{eq:Thm1Bound}
\mathcal{R}< & \underset{\mathcal{S},\overline{\mathcal{S}},\mathcal{S}_1,...,\mathcal{S}_p}{min} ~ \frac{1}{p} I(Y_R;\{X_{f,R}^k|k\in\mathcal{S}\}|\{X_{f,R}^k|k\in\overline{\mathcal{S}}\}, \nonumber \\
& \{X_{f,R}^k\oplus Z_i|k\in\mathcal{S}_i\} ~\forall~ i\in\{1,...,p\}).
\end{align}
For the special case when $\ell=2$, the set of bounds described by \eqref{eq:Thm1Bound} are equivalent to
\begin{align} \label{eq:QPSKRateBound}
\mathcal{R} < min \{&\frac{1}{2}I(Y_R;X_R^1,X_R^2),I(Y_R;X_R^1|X_R^2),I(Y_R;X_R^2|X_R^1), \nonumber \\
&I(Y_R;X_R^1,X_R^2|X_R^1\oplus Z_1,X_R^2\oplus Z_1)\}.
\end{align}

Note that
\[
I(Y_R;X_R^1,X_R^2|X_R^1\oplus Z_1,X_R^2\oplus Z_1)=I(Y_R;X_R^1,X_R^2|X_R^1\oplus X_R^2)\}.
\]
That is, $\{X_R^1\oplus Z_1,X_R^2\oplus Z_1\}$ and $\{X_R^1\oplus X_R^2\}$ carry the same information about $X_R^1$ and $X_R^2$.

\begin{figure*}[ht]
\subfigure[$\ell\mathcal{R}_f(h_A,h_B)$ vs. $\theta$ for each $f\in\mathcal{F}$.]{
\includegraphics[width=0.32\textwidth]{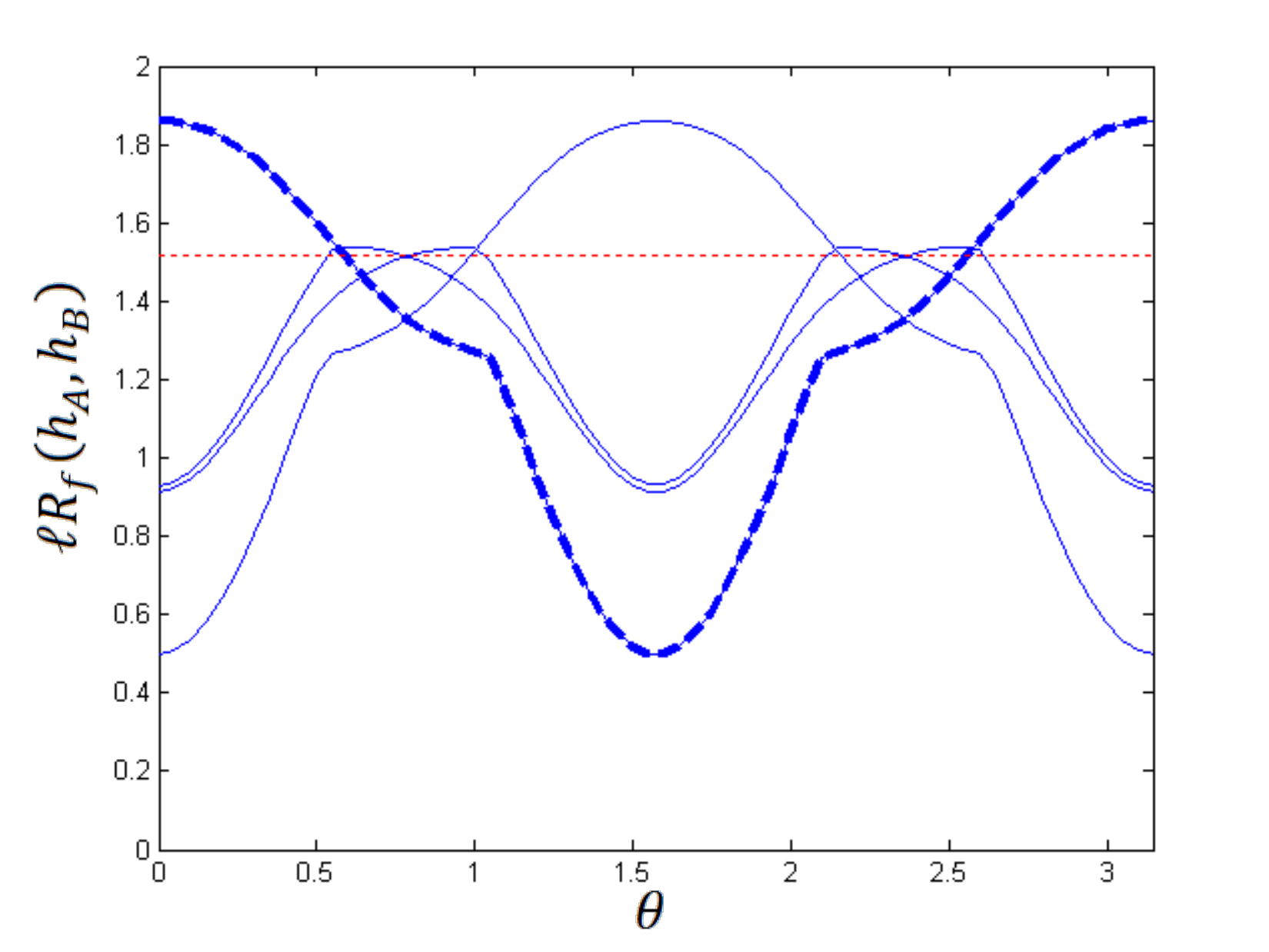}
\label{fig:Iyx_vs_theta_MLC}
}
\subfigure[$I(Y_R;f(X_A,X_B))$ vs. $\theta$ for each $f\in\mathcal{F}_{GF4}$.]{
\includegraphics[width=0.32\textwidth]{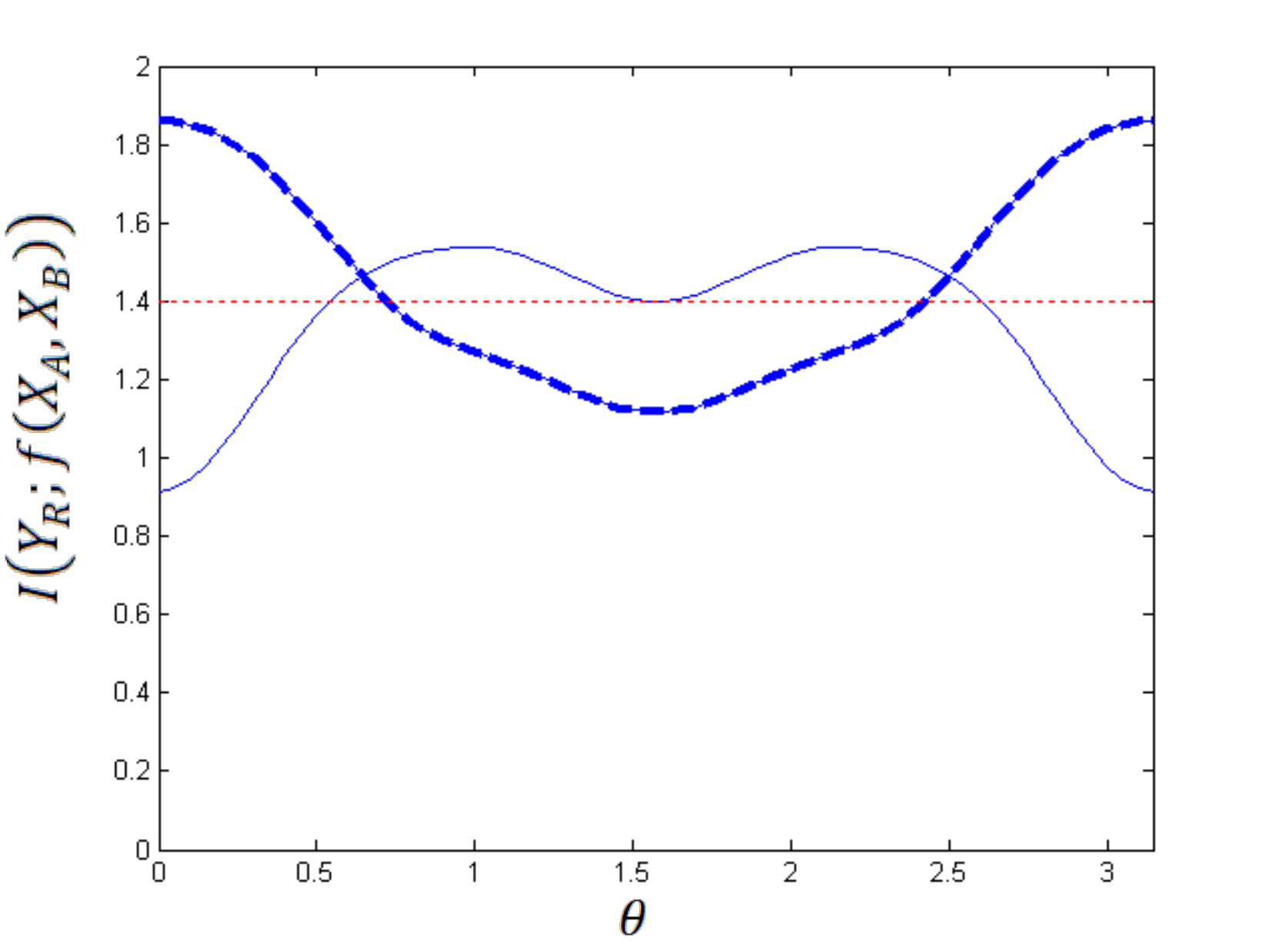}
\label{fig:Iyx_vs_theta_GF4}
}
\subfigure[Universally Achievable rates vs. SNR(dB).]{
\includegraphics[width=0.32\textwidth]{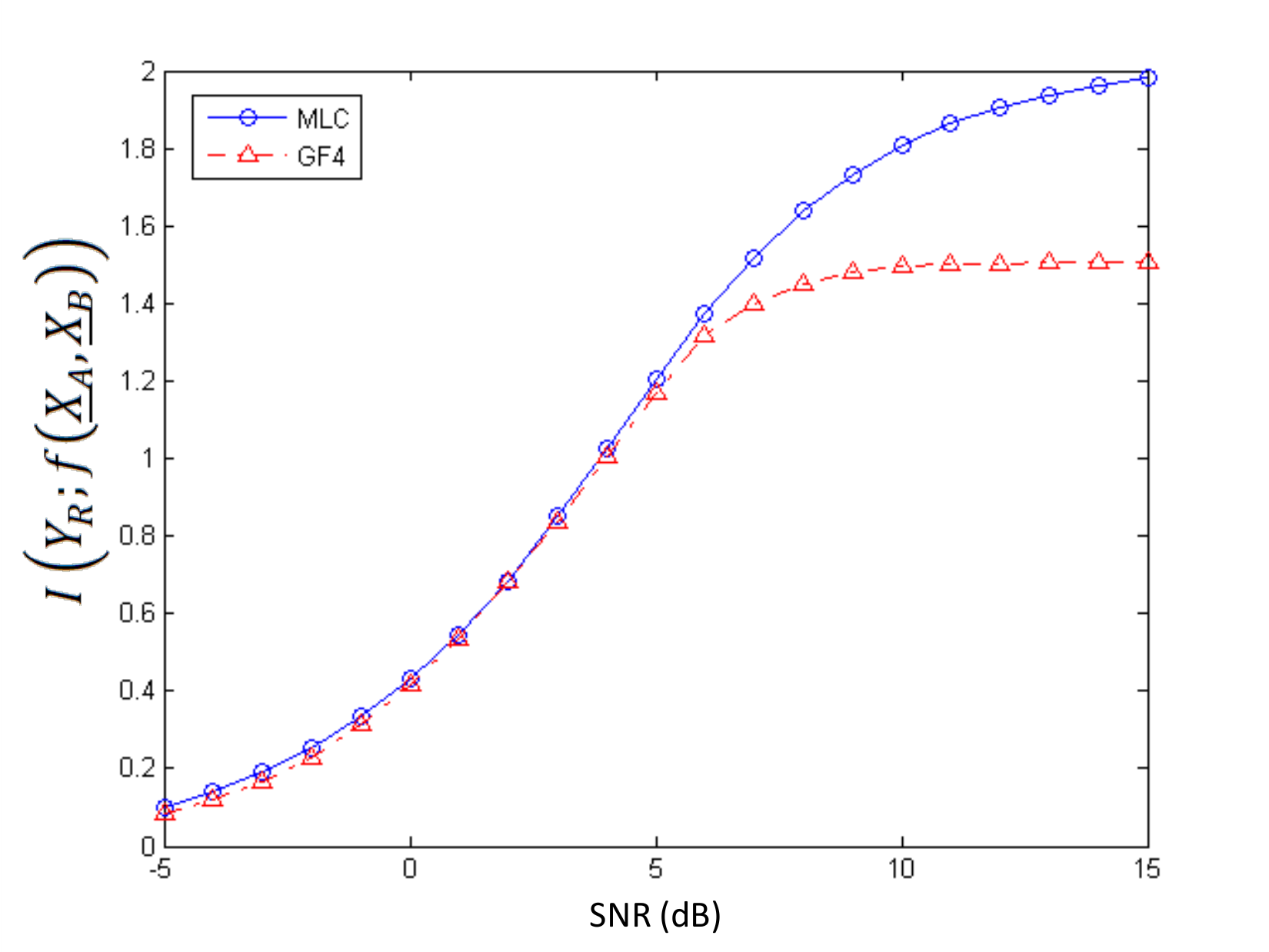}
\label{fig:Iyx_vs_snr}
}
\caption{Numerical Results}
\end{figure*}

\begin{proof}
The detailed proof is not included here because of space limitations. However, the key steps in the proof are outlined below.

Our proof uses the standard approach of deriving upper bounds on the probability of error for a joint typicality decoder averaged over a carefully chosen ensemble of codes. The ensemble considered here is the ensemble obtained by using random cosets of the {\em same linear} code for the different signalling levels in the multilevel coding scheme. The linear code is chosen from the ensemble of linear codes with randomly chosen entries in the generator matrix. The use of the same linear code in each level is an important ingredient in our proposed scheme since we allow the relay to freely take linear combinations of codewords from different signalling levels. However, this is also what complicates the proof. The ensemble used here is different from the often used ensemble of random coset codes used at each level in the multilevel coding scheme since the latter ensemble allows for independently chosen codes at each level. While the latter ensemble has been used widely to obtain achievable rates for MLC for the point to point channel and the multiple access channel, the former ensemble has not been analyzed in detail in the literature. The key contribution of our proof in the Appendix is to derive the achievable rates with the former ensemble with identical linear codes at each level.

This can be accomplished since the use of the same linear code at each level ensures that for each $f \in \mathcal{F}$, $\mathcal{C}_{f,R}^k,~k\in\{1,...,\ell\}$ is a member of the ensemble used at the transmitters. The main complication that arises from this is that the pairwise independence assertion that is required in typical channel coding proofs \cite{gallager1968information} does not hold for certain classes of error events. Particularly, it is possible for the relay to correctly decode some rows of $\mathbf{X}_{f,R}$ while others may be in error. We handle this by splitting the union bound for error probability into separate classes of error events which are conditionally pairwise independent.

%

The bound for the $\ell=2$ case can be derived by letting $\mathcal{S},\overline{\mathcal{S}},\mathcal{S}_1,\mathcal{S}_2\subseteq\{1,2\}$ take the following values respectively.
\begin{align}
&\{\mathcal{S}=\{1,2\},\overline{\mathcal{S}}=\emptyset,\mathcal{S}_1=\{1\},\mathcal{S}_2=\{2\}\} \nonumber \\
&\{\mathcal{S}=\{1\},\overline{\mathcal{S}}=\{2\},\mathcal{S}_1=\{1\}\} \nonumber \\
&\{\mathcal{S}=\{2\},\overline{\mathcal{S}}=\{1\},\mathcal{S}_1=\{2\}\} \nonumber \\
&\{\mathcal{S}=\{1,2\},\overline{\mathcal{S}}=\emptyset,\mathcal{S}_1=\{1,2\}\}.
\end{align}
Notice that the first three terms in $\eqref{eq:QPSKRateBound}$ are also required by the proof for the general MAC channel. The last bound is a result of the requirement that each signaling level uses a coset of the same linear code.
\end{proof}

\subsection{Universally Achievable Rate}
We say that a rate $\mathcal{R}$ is universally achievable over the set $\mathcal{H}\subset\mathbb{C}^2$ if there exists a fixed linear code $\mathcal{C}$ of rate $\mathcal{R}$ and coset matrices $\mathbf{\Lambda}_A$ and $\mathbf{\Lambda}_B$ such that for every $(h_A,h_B) \in \mathcal{H}$, the relay can reliably decode $\mathbf{X}_{f,R}$ for some $f\in\mathcal{F}$. That is some $\mathbf{X}_{f,R}$ can be decoded with arbitrarily small probability of error in the usual information-theoretic sense. The main result in this section is the following theorem.

\emph{Theorem 2:} For a fixed $f\in\mathcal{F}$ and $(h_A,h_B)$, define $\mathcal{R}_f(h_A,h_B)$ as the supremum of rates satisfying \eqref{eq:Thm1Bound} where $\underline{x}_{f,R}=f(\underline{x}_A,\underline{x}_B)$. For any finite set of channel gains, $\mathcal{H}\subset\mathbb{C}^2$, any rate $\mathcal{R}$ such that
\begin{equation} \label{eq:Thm2Bound}
\mathcal{R} < \underset{(h_A,h_B)\in\mathcal{H}}{min} ~ \underset{f\in\mathcal{F}}{max} ~ \mathcal{R}_f(h_A,h_B)
\end{equation}
is universally achievable.

\begin{proof}
The above result follows from the fact that the MLC scheme with transmission rate $\mathcal{R}$ is universal in the sense that the induced codebooks are simultaneously optimal for decoding any function $f \in \mathcal{F}$ if $\mathcal{R}_f(h_A,h_B) > \mathcal{R}$. This permits the relay to decode any function $f$ for which  $\mathcal{R} < \mathcal{R}_f(h_A,h_B)$ since $\mathcal{R}_f(h_A,h_B)$ corresponds to the information rate corresponding to the uniform i.i.d distribution. Hence, for every $(h_A,h_B)\in\mathcal{H}$ we can choose a function $f$ at the receiver such that $\mathcal{R}_f(h_A,h_B) > \mathcal{R}$ and the result follows. The proof for the existence of a single coset code $\{\mathcal{C},\mathbf{\Lambda}_A,\mathbf{\Lambda}_B\}$ which is good for the finite set $\mathcal{H}$ is omitted for space.
\end{proof}

Note that in order for this problem to be practically interesting, the set $\mathcal{H}$ should be meaningfully defined.

\section{Numerical Results} \label{sec:NumericalResults}

\subsection{Numerical Results for QPSK}
As an example, consider the case where nodes A and B transmit symbols from a QPSK constellation with Gray Labeling. Fig.~\ref{fig:Iyx_vs_theta_MLC} shows a plot of the achievable information rate $\ell \mathcal{R}_f(h_A,h_B)$ as given in \eqref{eq:QPSKRateBound} for each function $f \in \mathcal{F}$ dependent on the phase difference $\theta = \theta_A-\theta_B$ for an SNR of $7~dB$. $\mathcal{H}$ is the set of channel gains
\begin{equation} \label{eq:HSetDef}
\mathcal{H} = \{(h_A,h_B)|h_A=e^{j\theta_A},h_B=e^{j\theta_B}\}
\end{equation}
where $\theta_A,\theta_B\in\{0,\frac{\pi}{m},...,2\pi\}$ for a finite integer $m$. Thus $|\mathcal{H}|$ is finite but approximates a the selection of any value of $\theta_A$ and $\theta_B$ arbitrarily closely. The dotted line indicates the universally achievable rate in bits per complex symbol for the proposed scheme which satisfies Theorem 2 for $\mathcal{H}$. Note that different functions provide the best performance for different values of $\theta$ which reiterates the substantial benefit of decoding adaptively.
Notice that a small increase in rate makes reliable decoding impossible for any $f\in\mathcal{F}$ for a significant range of $\theta$; however, there are many $(h_A,h_B)\not\in\mathcal{H}$ such that $\exists f\in\mathcal{F}$ for which reliable decoding is possible.

\subsection{Coding over $GF(4)$}
It is interesting to use this QPSK example to compare our MLC scheme to the case where nodes A and B encode their a data in $\mathbb{F}_4$ using a linear code $\mathcal{C}_{GF4}$ of rate $\mathcal{R}_{GF4}$. The relay uses the set of decoding functions $\mathcal{F}_{GF4}$ corresponding to linear combinations of codewords in $\mathbb{F}_4$ of the form
\begin{equation}
\underline{v}_R = f(\underline{v}_A,\underline{v}_B) = \alpha \underline{v}_A \oplus \beta \underline{v}_B ,~ \alpha,\beta\in \mathbb{F}_4\backslash\{0\}.
\end{equation}
The value of $I(Y_R;f(X_A,X_B))$ for each possible $f\in\mathcal{F}_{GF4}$ is plotted as a function of $\theta$ in Fig. \ref{fig:Iyx_vs_theta_GF4} with an SNR of $7~dB$. Again the dotted line represents universally achievable rate for the $\mathcal{H}$ in \eqref{eq:HSetDef}.

\subsection{Comparison of Proposed Techniques}
%

These numerical results illustrate that the proposed MLC scheme facilitates better decoding flexibility at the relay than coding over $\mathbb{F}_4$ for this example. In fact, an analysis of these functions based on the labeling of points in $\mathcal{Q}_{R}$ shows that $\mathcal{F}_{GF4}\subset\mathcal{F}$. However, this improved flexibility comes at the cost of additional rate constraints on each $f\in\mathcal{F}$. The thick dashed line in Figs. \ref{fig:Iyx_vs_theta_MLC} and \ref{fig:Iyx_vs_theta_GF4} represents the rate which is achievable if the relay decodes using some $f$ which is equivalent to the componentwise xor operation for multilevel coding or finite field addition for $\mathbb{F}_4$. The difference between these curves illustrates the effects of the additional rate constraints imposed by \eqref{eq:Thm1Bound}. In Fig. \ref{fig:Iyx_vs_theta_MLC} the last term $I(Y_R;X_R^1,X_R^2|X_R^1\oplus Z_1,X_R^2\oplus Z_1)$ in \eqref{eq:QPSKRateBound} is dominant if $\theta\approx\frac{\pi}{2}$ for determining the achievable rate for this function. In Fig. \ref{fig:Iyx_vs_theta_GF4} we see that this term does not need to be satisfied if nodes A and B use a linear code in $\mathbb{F}_4$.

The universally achievable rate for the $\mathcal{H}$ in \eqref{eq:HSetDef} (i.e. the constant value given by the dotted line in Figs. \ref{fig:Iyx_vs_theta_MLC} and \ref{fig:Iyx_vs_theta_GF4}) is plotted as a function of SNR in Fig. \ref{fig:Iyx_vs_snr} for the cases where the relay uses $\mathcal{F}$ or $\mathcal{F}_{GF4}$. This value asymptotically approaches 1.5 bits per symbol for coding over $\mathbb{F}_4$. From Fig. \ref{fig:Iyx_vs_theta_GF4}, this appears to occur because $\mathcal{F}_{GF4}$ does not provide the relay with a decoding function which works well when $\theta\approx\frac{\pi}{2}$. This represents an extreme case, because the event $|h_A|=|h_B|$ occurs with probability zero for many random fading processes. However, this illustrates that for PLNC it is possible for the universally achievable rate to be limited by specific $(h_A,h_B)\in\mathcal{H}$ even if each $|h_A|$ and $|h_B|$ is large. While the proposed scheme is better than choosing only one decoding function, it still does not entirely eliminate the interference that results from the signal constellations not aligning perfectly at the receiver. Hence, the achievable rate in the high SNR regime will be limited by this interference in addition to the limitation imposed by a finite-sized constellation.



\section{Concluding Remarks} \label{sec:ConcludingRemarks}
In this paper, we have proposed a coding scheme based on MLC for compute and forward or PLNC for the case when the channel is perfectly estimated at each receiver but unknown to each transmitter. We showed that MLC allows for decoding of a set of functions of the transmitted messages and the relay can choose one function from this set depending on the channel coefficients. In Theorem~1, we obtained an achievable rate for a fixed decoding function and channel realizations. In Theorem~2, we obtained a numerically computable expression for the universally achievable information rate over a set of channel realizations. Numerical results for QPSK suggest that the proposed scheme significantly outperforms the use of a fixed decoding function with binary linear codes and is better than using linear codes over $\mathbb{F}_4$.

\bibliographystyle{ieeetr}
\bibliography{lattice}

\end{document}